\documentclass[12pt]{article}
\usepackage{times}
\usepackage{geometry}
\usepackage{graphicx}
\usepackage[dvipsnames]{xcolor}
\usepackage{natbib}
\usepackage{enumitem,amssymb}
\usepackage{ragged2e}
\newlist{thematic}{itemize}{8}
\setlist[thematic]{label=$\square$}
\usepackage{pifont}

\usepackage{hyperref}
\newcommand\myshade{100}
\colorlet{mylinkcolor}{violet}
\colorlet{mycitecolor}{YellowOrange}
\colorlet{myurlcolor}{Aquamarine}
\hypersetup{
  linkcolor  = mylinkcolor!\myshade!black,
  citecolor  = mycitecolor!\myshade!black,
  urlcolor   = myurlcolor!\myshade!black,
  colorlinks = true,
}
\usepackage[capitalize]{cleveref}

\begin{document}

\title{International Astronomical Consortium for High-energy Calibration
\\
{\normalsize IACHEC 2020/2021 Pandemic Report}
\\
\bf{Summary of the Virtual IACHEC Meetings}
}

\author{
K.~K.~Madsen$^1$,
V.~Burwitz$^2$,
K.~Forster$^3$
C.~E.~Grant$^4$,
M.~Guainazzi$^5$, \and
V.~Kashyap$^6$, 
H.~L.~Marshall$^4$, 
E.~D.~Miller$^4$, 
L.~Natalucci$^7$, 
P.~P.~Plucinsky$^6$, \and
Y.~Terada$^{8,9}$
}

\date{\today}   

\newcommand{\artxc}{\textit{ART-XC}}
\newcommand{\erosita}{\textit{eROSITA}}
\newcommand{\rosat}{\textit{ROSAT}}
\newcommand{\xmm}{XMM-\textit{Newton}}
\newcommand{\chandra}{\textit{Chandra}}
\newcommand{\suzaku}{\textit{Suzaku}}
\newcommand{\swift}{\textit{Swift}}
\newcommand{\nicer}{\textit{NICER}}
\newcommand{\astrosat}{\textit{Astrosat}}
\newcommand{\nustar}{\textit{NuSTAR}}
\newcommand{\hxmt}{\textit{Insight-HXMT}}
\newcommand{\hitomi}{\textit{Hitomi}}
\newcommand{\xrism}{\textit{XRISM}}
\newcommand{\integral}{\textit{INTEGRAL}}
\newcommand{\fermi}{\textit{Fermi}}
\newcommand{\athena}{\textit{Athena}}
\newcommand{\ep}{\textit{Einstein Probe}}
\DeclareRobustCommand{\ion}[2]{\textup{#1\,\textsc{\lowercase{#2}}}}
\newcommand{\cstat}{{\tt c-stat}}

\maketitle

{\centering
  $^1$CRESST and X-ray Astrophysics Laboratory, NASA Goddard Space Flight Center, Greenbelt, MD 20771, USA \\
  $^2$Max Planck Institute for Extraterrestrial Physics, Germany\\
  $^3$Cahill Center for Astronomy and Astrophysics, California Institute of Technology, USA \\
  $^4$Kavli Institute for Astrophysics and Space Research, Massachusetts Institute of Technology, USA \\
  $^5$ESA-ESTEC, The Netherlands \\
  $^6$Center for Astrophysics $|$ Harvard \& Smithsonian (CfA), USA \\
  $^7$IAPS-INAF, Italy \\
  $^8$Saitama University, Japan \\
  $^9$Japan Aerospace Exploration Agency, Institute of Space and Astronautical Science, Japan
}

\vspace{5mm}
\begin{center}
{\bf \large Abstract} \\
\end{center}

In this report we summarize the activities of the International Astronomical Consortium for High Energy Calibration (IACHEC) and the work done since the last in-person meeting in Japan (Shonan Village Center), May 2019, through two virtual meetings that were held in November 2020 and May 2021. The on-line only meetings divided the contents of the usual in-person workshop between mission updates and working group updates. The November meeting was dedicated to mission calibration updates and the current status of the cross-calibration between \nustar, \swift, and \nicer, which frequently join together in observations of bright transients, and a review of the \xmm\ and \chandra\ cross-calibration. Results between \nustar\ and \swift\ overall show good agreement, but issues persist in the overlap region 3--5~keV for bright source with large dust scattering halos. The \nicer\ cross-calibration is still progressing and evolving, while for the \xmm\ and \chandra\ cross-calibration systematic differences both in the absolute flux and spectral shape determination still exists on different classes of sources. The meeting in May was focused on the Working Group progress and reports summarized here.

\section{Introduction}

The International Astronomical Consortium for High Energy Calibration (IACHEC)\footnote{\href{http://iachec.org}{\tt http://iachec.org}} is a group dedicated to supporting the cross-calibration of the scientific payload of high energy astrophysics missions with the ultimate goal of maximizing their scientific return. Its members are drawn from instrument teams, international and national space agencies, and other scientists with an interest in calibration. Representatives of over a dozen current and future missions regularly contribute to the IACHEC activities. 

IACHEC members cooperate within Working Groups (WGs) to define calibration standards and procedures. The objective of these groups is primarily a practical one: a set of data and results produced from a coordinated and standardized analysis of high-energy reference sources that are in the end published in refereed journals. Past, present, and future high-energy missions can use these results as a calibration reference.

Table \ref{table:WGoverview} summarizes the WGs active during this report and their primary projects and areas of responsibility.

\begin{table}[]
    \centering
    \begin{tabular}{| p{0.30\linewidth}|p{0.2\linewidth}|p{0.5\linewidth}|}
        \hline
         Working Group & WG Chair & Projects  \\
         \hline
         \hline
         Calibration Statistics & Vinay Kashyap & Quantifying response uncertainties; statistical methods; Concordance project \\
         \hline
         Clusters of Galaxies & Eric Miller & Multi Mission Study of selected targets from the HiFLUGCS sample\\
         \hline
         Contamination & Herman Marshall & Definition, measurement, and mitigation of molecular contaminant \\
         \hline
         Coordinated Observations & Karl Forster & Organization of yearly cross-calibration campaign of 3c273; investigation of potential cross-calibration candidate 1ES 0229+200; publication of 3c273 cross-calibration campaign\\
         \hline
         Detectors and Background & Catherine Grant & Forum for discussion of detector effect; background modeling\\
         \hline
         Heritage & Matteo Guianazzi & Curation of the IACHEC work; the IACHEC source database (\href{http://iachecdb.iaps.inaf.it}{\tt ISD})\\
         \hline
         Non-thermal SNR & Lorenzo Natalucci & Cross-calibration with the Crab and G21.5-0.3\\
         \hline
         Thermal SNR & Paul Plucinsky & Cross-calibration with 1E 0102.2$-$7219 (E0102) and N132D; definition of standard models \\
         \hline
         Timing & Yukikatsu Terada & Timing calibration across missions, see Table \ref{table:timing} \\
         \hline
         White Dwarf and Isolated Neutron Stars & Vadim Burwitz & Cross-calibration with RX\,J1856.5$-$3754 and 1RXS\,J214303.7$+$065419 \\
         \hline
    \end{tabular}
    \caption{IACHEC Active Working Groups}
    \label{table:WGoverview}
\end{table}

\section{Virtual Meeting Summary} 
In 2020 and 2021, the IACHEC suspended all in-person meetings due to the Covid-19 pandemic and organized instead two virtual meetings which split the usual 3.5 days into two sections: mission updates and WG activity and progress. 

The first part was held on November 23-24, 2020, and progressed over two days of three hours each. The first day focused on observatory calibration updates from \astrosat, \chandra, \integral, \nicer, \nustar, \swift, and \xmm. The first half of the second day reported on the status of the cross-calibration between \nicer, \swift, and \nustar, and the continued cross-calibration of \chandra\ and \xmm. The cross-calibration between \nustar\ and \swift\ is largely understood and good for most sources up to an intermediate brightness and N$_H < 1 \times 10^{22}$, but issues persist for very bright sources, which typically also have dust scattering halos, and has to do with how the angular dependent dust scattering halo gets included into the extractions. No recommended solution to the problem exists yet. For \chandra\ and \xmm, Michael Smith presented results from cross-calibration with blazars 3c273, PKS 2155-304, and H1426+428, which despite clear improvements in the last decade (compare, for instance, Nevalainen et al. 2010 vs. Madsen et al. 2017), exhibits systematic differences both in the absolute flux (see Section \ref{s:t_snr}) and the spectral shape determination, whose origin remains still not fully ascertained and remains a subject of investigation. The second half of the day had a plenary on pileup given by Richard Saxto and a discussion on statistical best practices, focused on likelihood definitions by G. Belanger and background modelling led by Eric Miller, as well as updates on the Concordance project (Marshall et al., 2021). All talks were recorded and made available at the IACHEC web-page \footnote{\href{https://iachec.org/2020-iachec-online-meeting-date/}{\tt IACHEC 2020 Symposium}}. Both days were attended by around 70 people.

The second meeting was held on May 17-19, 2021, and focused on the WGs presenting their past activities and future plans, which are summarized here in this report. On separate days from the WG meeting, the IACHEC also hosted a series of three plenaries on the topic of: '\textit{Planning in-flight calibration for XRISM}', '\textit{First Results and Calibration of eROSITA}', and '\textit{The Calibration of IXPE}'. The WG meeting and plenaries were recorded and made available on the IAHCEC web-page\footnote{\href{https://iachec.org/2021-virtual-meetings/}{\tt IACHEC 2021 Spring WG meeting}}. Again, the meeting and plenaries were attended by around 70 people all days.

A couple of the high level topics that did not fall into a specific WG category came out of these meetings and we report on them here:
\begin{itemize}
    \item It was commented that the work IACHEC has done and documented on the web-pages has been immensely useful, and it was requested that the IAHCEC continue to update the information.
    \item The IACHEC held a discussion session with the community on how to improve our work, and it was reported that in general the papers the IACHEC have published have been very useful and easy to use, but that it isn't easy to search for known calibration issues across observatories. A centralized place to search for such things would be desirable. 
    \item Along the same lines, data analysis issues that could be cross-calibration related are not always reported to helpdesks of both of the involved observatories, and the picture of cross-calibration problem may therefore not be complete.
    \item There is an increasing interest in background modeling, and since presently the effort of building observatory specific model backgrounds has been scattered throughout the WG, it was discussed to unify it into a single WG. 
\end{itemize}

Finally, G. Belanger has stepped down as the Communications WG chair, and in the upcoming meetings the IACHEC will be discussing the future scope of the Communications WG.

Publications out of the IACHEC effort this year comes from the Calibration Working group, who have submitted a work on Concordance (discussed in Section \ref{s:calstat}). At the time of writing the work, Marshall et al. (2021), has been accepted for publication in the Astrophysical Journal.

\section{Working Group reports}

\subsection{Calibration Uncertainties: CalStats}\label{s:calstat}

The CalStats WG focuses on statistical methodology as applied to calibration data and analysis. This includes documenting and recommending good analysis practices, developing techniques that are mathematically robust, and finding methods that deal with commonly encountered calibration analysis problems.
Members of this WG held two group meetings (on 2020-May-5 and \href{https://iachec.org/calibration-statistics/#2020dec1}{2020-Dec-01}), in addition to organizing a special session at AAS 238 (\href{https://hea-www.harvard.edu/AstroStat/aas238/}{Unaccounted Uncertainties: The Role of Systematics in Astrophysics} 2021-Jun-7), participating in plenary sessions during the Virtual Symposium in 2020~Nov (\href{https://iachec.org/2020-iachec-online-meeting-date/}{Statistical Best Practices}) and the \href{https://iachec.org/2021-virtual-meetings/}{Spring WG meeting} in 2021~May, and hosting several talks relevant to the group's aims (\href{https://astrostat.org/jsm2020/index.html#session-431}{K.Madsen 2020}, at JSM; \href{https://hea-www.harvard.edu/AstroStat/CHASC_2021/index.html#hm+yc_20200908}{Marshall \& Chen 2020}, at CHASC; \href{https://hea-www.harvard.edu/AstroStat/CHASC_2021/index.html#dj_20201208}{D.Jerius 2020}, at CHASC; \href{https://hea-www.harvard.edu/AstroStat/aas238/#schedule}{Chen, Y. 2021}, at AAS).  Some specific highlights are listed below:

\paragraph{Concordance:} Considerable progress has been made in the Concordance project since the first publication of the method (Chen et al.\ 2019).  The technique has been updated to account for correlations that persist across energy in instrument effective areas, several more example analyses have been performed, and the method has been explicitly validated with simulations (Marshall et al.\ 2021).

\paragraph{Systematic Uncertainties:} Efforts are underway to construct an effective area uncertainty sample (see Drake et al.\ 2006) for AstroSat instruments.  Once constructed, such a sample can be included within standard spectral analysis using the pyBLoCXS package (van Dyk et al.\ 2001).  A similar process to account for uncertainties in atomic data emissivities has also been developed (Yu et al.\ 2018) and applied to Chandra gratings data on Capella (Yu 2021, Yu et al.\ 2021 in preparation).

\paragraph{Background:} The CalStats WG coordinated with the Detectors and Background WG (see Section~\ref{s:det_bkg}) to explore different aspects of background model development and use.  Several independent efforts to characterize the backgrounds in different instruments -- Suzaku (E.Miller), Chandra/ACIS (T.Gaetz), Chandra/HRC (B.Nevin \& G.Tremblay), Chandra/LETGS+HRC-S (B.Wargelin), eROSITA (K.Dennerl) -- are underway.  Our goal is to ultimately provide easy-to-use scripts that can be used in spectral modeling environments like Sherpa and XSPEC.  In addition, sophisticated analysis techniques that can characterize the departure of observed data from the nominal model are also being developed (Algeri 2020).

\paragraph{Polarization:} H.\ Marshall described recent efforts to develop a coherent statistical framework to describe and model polarimetry data in the context of new missions like IXPE (see Marshall 2021a,b).

\paragraph{Best practices:} G.Belanger presented two talks (one at the plenary, and one at the WG meeting) describing how a proper likelihood should be defined for timing data (see Belanger 2013,2016).
M.Bachetti presented several tools -- \href{{https://github.com/nanograv/pint}}{\tt PINT},
\href{https://github.com/stingraysoftware/stingray}{\tt Stingray}, and 
\href{https://github.com/stingraysoftware/hendrics}{\tt hendrics}
-- to assess timing calibration.
V.Kashyap contributed a chapter on the basics of Astrostatistics to the Tutorial Guide to X-ray and Gamma-Ray Astronomy (Kashyap 2020).
V.Kashyap has also provided Sherpa/python and IDL scripts that implement the {\tt cstat} goodness-of-fit measure of Kaastra (2017).  K.Arnaud has implemented a method in XSPEC to encode high-resolution information in RMFs (based on Kaastra \& Bleeker 2016) that can be used for high-res spectral analysis for data obtained with upcoming missions like XRISM and ARCUS.


\subsection{Clusters of Galaxies}\label{s:clusters}

This WG leverages clusters of galaxies as cross-calibration standard X-ray sources. These systems have several advantages compared to other sources: the hot, X-ray-emitting intracluster medium is very stable in flux, it is X-ray bright across a broad band, and hotter clusters have a fairly simple continuum-dominated spectrum. However, they are spatially extended and often contain bright, variable AGN, which complicates comparison between instruments with very different imaging characteristic. The WG, while very active throughout the history of the IACHEC (Nevalainen et al.\ 2010, Kettula et al.\ 2013, Schellenberger et al.\ 2015), had been dormant for two years before meeting virtually in May 2021. It was recognized that the WG is valuable, and that recently launched and upcoming missions including eROSITA, XRISM, and ATHENA have extensive cluster science goals and will use clusters as calibration targets, and that the WG should continue.

Much of the discussion focused on the status of the Multi-Mission Study (MMS), a project aiming to compare X-ray spectroscopic results of a sample of clusters obtained with on-going and past X-ray missions and instruments. Given the intervening time, additional instruments can now be included; we identified WG members to represent XMM-Newton EPIC MOS and pn, Chandra ACIS, Suzaku XIS, Swift XRT, NuSTAR, and HXMT, and we hope to include ROSAT PSPC, eROSITA, ASTROSAT, and NICER. The representative for each instrument will identify and gather existing data, apply the most recent calibration, extract spectra and responses, and provide these to the WG chair for MMS cross analysis. This data will also be provided to the CalStats WG for inclusion in the Concordance effort (see Section \ref{s:calstat}). We expect to select about four clusters for an initial comparison, drawing from the HiFLUGCS sample (Reiprich \& B\"ohringer 2002) hot ($kT>6$ keV), nearby ($z<0.1$) systems with at least 100,000 counts accumulated in the central 6$^\prime$ in each instrument, observed no more than 3$\prime$ off-axis.

I.~Valtchanov alerted the WG to a recent update to the chip gap/bad pixel correction on XMM-Newton EPIC pn (Nevalainen et al. 2021) that solves a previously reported problem with scaling spectra obtained with different instruments. In the EPIC instruments, a significant fraction of the FOV is obscured by the CCD gaps. Until this update, the flux was scaled linearly with obscured area fraction to correct for lost photons, technically only correct for spatially uniform emission. The update uses the EPIC MOS image to correct for the obscured areas on EPIC pn, reducing the residual error in the ARF to 0.1\%.

\subsection{Contamination}\label{s:contam}

The Contamination WG shares information about soft X-ray instruments that suffer from molecular contamination (e.g., Marshall et al.\ 2004, Koyama et al.\ 2007, O'Dell et al.\ 2013).  Since its inception, the WG has covered three broad topics: (1) measurements of contamination on various instruments and missions; (2) mitigation for current instruments; and (3) contamination prevention for future instruments.  The WG met for one session which included updates from several operating missions along with participation from calibration scientists of past and future missions.

Michael Smith (ESAC) reported that the XMM pn camera still shows no sign of contamination, while
the contamination on the MOS and RGS detectors is growing mildly.  Measurements of the isolated
neutron star RX J1856-3754 indicate that there may be 20-30 nm more contaminant on the RGS
over the past 2-3 years than predicted in the current calibration file.  The XMM MOS1
instrument still has less than 20 nm of contaminant, which is marginally detectable, while
MOS2 is mildly thickening at the rate predicted in the current calibration file.

Herman Marshall and Akos Bogdan showed results for the Chandra ACIS instrument that
are currently modeled by linear growth at about 160 nm/yr at the center of the detector
and at a $\sim$ 50\% higher rate near the edges.  The total depth is about a factor of 10
thicker than on the XMM MOS2 or RGS.

Besides the XMM pn camera, instruments that have shown no or very little contaminant include
the Swift XRT, Hitomi, and (so far) eROSITA.  Vadim Burwitz reported
that the eROSITA operations involved 6 days of
outgassing each component on the ground at $+80$ C and minimizing space around filters
for contaminants to reach the detectors.  Even cables were outgassed for extended periods
and stored in clean environments.  Eric Miller reported
that the XRISM CCDs will be protected in a similar fashion
to eROSITA, with filters warmed to $+30$ C that are 20 cm from the CCDs.
The plans for Athena are similar to that of eROSITA for the WFI and to XRISM/Resolve
for the XIFU.

Herman Marshall presented a plan for a white paper that could evolve into a
refereed publication.  The plan was based on previous suggestions to start with
an overview of objectives (to compare instruments and mitigation methods),
report status for each of 7-10 instruments, and present mitigation/monitoring
plans for upcoming missions which include Athena, Arcus, XRISM, IXPE, and SMILE/SXI.
A template for the paper will be presented at the next meeting so that members
may fill out their sections.

\subsection{Coordinated Observations}\label{s:coord}

The objectives of the IACHEC Coordinated Observations WG are to coordinate new observations jointly among different telescopes, analyze those observations, and publish the results.  For the 2019 IACHEC meeting, there were two presentations on cross-calibration projects and significant discussion of potential new observations.

Communications between the calibration and operations teams for the high-energy missions continued during the pandemic and a number of coordinated calibration observations were organized for 2019-2021. These included the annual multi-mission cross-calibration campaign of observations of the quasar 3C 273 as well as joint INTEGRAL, NuSTAR, and XMM-Newton observations of the Crab nebula.

The campaign of observations of the quasar 3C 273 involving Chandra, INTEGRAL, NICER, NuSTAR, Gehrels-Swift, and XMM-Newton observatories were organized and performed on 2019-07-02 to 03, 2020-07-06 to 07, and 2021-06-09 to 10. Analysis of the data from these observations, along with those performed each year from 2015, is underway with the goal of updating the study of cross-calibration results based on 2012 and 2013 observations of 3C 273 and PKS 2155-304 (Madsen et al. 2017a).

The coordination of these observations was aided by new channels for communicating the status of calibration observations, initiated in the last year. These include a dedicated slack channel for the working group, an email list server hosted at Caltech (iachec-co-obs \_at\_ lists.srl.caltech.edu), and a set of Google-sheets that present detailed information about IACHEC observatories, calibration targets, and observation scheduling to aid in coordination between observatories\footnote{For more information contact iachec-co-obs \_at\_ lists.srl.caltech.edu}. 

A meeting of the coordinated observations working group (COWG) was held over zoom on May 13th, 2021 with 16 members in attendance. The agenda for the meeting included updates on the annual observing campaigns and a discussion about the additional avenues for communications within the COWG. 

Information about the recently adopted IVOA standards for target visibility and observation scheduling\footnote{Details about the standards can be found at https://www.cosmos.esa.int/web/vovisobs\_protocols/home} was presented by Karl Forster. INTEGRAL, XMM-Newton, and Chanda operations have already adopted these standards and there is a demonstration web client (TOBY\footnote{http://integral.esa.int/toby/}) that can display observing schedules and target visibility for multiple observatories. Adoption of these standards is being planned by other high energy missions as well as a number of ground based observatories. This will become invaluable for coordination of future cross-calibration campaigns.

Details about the 2021 cross-calibration campaign of observations of 3C 273 were discussed at the meeting with a preliminary schedule presented based on the visibility constraints for the Chandra observation. The Gehrels-Swift and NICER observatories planned to join the campaign, along with the INTEGRAL, XMM-Newton and NuSTAR observatories. 

Unfortunately, due to UVIT telescope operations constraints, it is not possible for Astrosat to observe 3C 273, and so an alternative astrophysical calibration target has been investigated. Norbert Schartel and Felix Fuerst presented an analysis of the potential of the blazar 1ES 0229+200 (J2000 38.20250 +20.288333) as an astrophysical calibration source. Blazars can provide good calibration sources as they have simple power-law spectra between 200 eV and 15 keV. The annual calibration target 3C 273 is a blazar but bright enough to cause pile-up issues in the sensitive soft X-ray instruments on Chandra, XMM-Newton and Gehrels-Swift. The investigation determined that the blazar 1ES 0229+200 should be bright enough to provide spectra of sufficient quality in reasonable exposure times, but faint enough to avoid pile-up issues.

This blazar also has the benefit of a very simple power-law spectrum extending to lower energies and so modeling of the spectrum can include data obtained simultaneously from the UV sensitive instruments on XMM-Newton (OM), Gehrels-Swift (UVOT), and Astrosat (UVIT). Although the X-ray flux from 1ES 0229+200 is expected to be about 20\% of that from 3C 273, the historical light-curve indicated significant flux variability\footnote{https://www.swift.psu.edu/monitoring/source.php?source=1ES0229+200}, though no spectral changes, so simultaneous observations will be required for cross-calibration investigations.

Plans for observations of 1ES 0229+200 by XMM-Newton and NuSTAR in August 2020 were presented at the meeting including coordinated observations by the TeV observatories H.E.S.S, MAGIC, and VERITAS and radio observations from the Effelsberg 100m observatory. The Astrosat, NICER, and Gehrels-Swift teams planned to investigate the possibility of joining the observing campaign.

The members of the working group meeting also discussed supporting calibration activities of the new missions that are planned to be launched in the next few years, IXPE\footnote{https://ixpe.msfc.nasa.gov} (December 2021), and XRISM\footnote{https://xrism.isas.jaxa.jp/en/} (2022-2023) and agreed to hold further meetings of the working group approximately 3 months before each launch to determine a schedule of supporting observations.

\subsection{Detectors and Background}\label{s:det_bkg}

The Detectors WG provides a forum for cross-mission discussion and comparison of detector-specific modeling and calibration issues, while the Background WG provides the same for measuring and modeling instrument backgrounds in the spatial, spectral and temporal dimensions.  In a typical year, the WG holds multiple sessions during the in-person meeting, covering a wide variety of topics and missions. These are generally well attended and can provide an opportunity for more junior members to present.  In the pandemic year, that function of the WG has gone dormant, however there have been relevant presentations at both the Fall 2020 and Spring 2021 IACHEC virtual meetings. 

The 2020 Online Symposium included a talk on best-practices for background modeling by Eric Miller, using observations of the supernova remnant N132D by \suzaku\ as an example case. Spring 2021 included a plenary session on \erosita\ calibration and results by Konrad Dennerl. The \erosita\ detector presents a difficult calibration problem, with 384 readout nodes for each of seven CCDs, which each need separate gain and CTI measurements. He discussed the substantial ground calibration program and the continuation of that program on-orbit.  He also discussed the particle background at L2 as measured by \erosita, the first X-ray observatory at that location. The background is higher than pre-launch simulations, but is much less variable than that seen in \xmm. Recordings and slides from the virtual presentations are available on the IACHEC web page.

\subsection{Heritage}\label{s:heritage}

The Heritage WG has the scope of preserving the IACHEC corpus of knowledge, know-how, and best-practices for the benefit of future missions and the community at large. The efforts of the Working Group are currently concentrated on ensuring that sufficient resources are allocated to create an "IACHEC Source Database" as the single repository of all data, instrument responses and data reduction and analysis software used in analyzing the calibration data published in IACHEC papers. An initial version of such a database was developed under a generous funding of the AHEAD initiative. However, this source of funding is no longer available for further development and long-term preservation.

\subsection{High-Resolution Spectra}\label{s:hires}

The High-Resolution Spectra Working Group (HRWG) aims to improve interpretation of spectra from high-resolution instruments. This year, two linelist projects were presented which will improve the modeling of high resolution spectra. Liyi Gu presented his work (Gu et al., 2020) on the Capella spectrum. This involved fitting over 500ks of \textit{Chandra} HETG data using existing databases and new Fe L-shell calculations made for the SPEX project. This greatly improved the fit in the Fe L-shell regions. In addition, by freeing all ion populations and allowing for ``correction factors'' to each line which still didn't fit, it was possible to estimate where emissivity calculations were an issue and where wavelengths were the bigger problem. This is a promising technique for producing line lists with both wavelength and intensity uncertainties attached. Norbert Schulz presented preliminary results from \textit{Chandra} ACIS/LETG observations in progress of $\theta^1$ Ori C. This has identified 110 line-like features, which are still undergoing identification. Both of these projects promise to provide an excellent source of line centroids to update spectral databases.

\subsection{Non-Thermal SNRs}\label{s:nt_snr}

The mission of the non-thermal SNR WG is to define the model of the two Pulsar Wind Nebula (PWN) standard candles:  G21.5-0.9 (below 10 keV) and of the Crab (above 10 keV). In the period of the pandemic, the WG had one virtual meeting in addition to  the 2021 Spring IACHEC WG meeting where they presented progress and updates. 

G21.5-0.9 is a plerionic pulsar wind nebula of a few mCrab, 3 arminutes across, and with no detectable pulsations, which makes it well suited as a calibration candle for a variety of instruments. It has been observed with almost all current X-ray observatories, and typically has been fit with a power-law spectrum between 0.1--10~keV. However, there is a hint of curvature or a break at $\sim$7~keV seen in both \nustar\ and \hitomi. Since the IACHEC paper by Tsujimoto et al. (2011), which investigated the cross-calibration of G21.5-0.9 between \chandra, \integral, {\em RXTE}, \suzaku, \swift, and \xmm, there has not been a concerted effort to compare observations with newer observatories, and seen in the light of the possible curvature or break in the bandpass, it was decided to start up a new investigation into defining a new standard spectrum for G21.5-0.9, with the goal to publish the results.  

For the Crab, L. Natulucci has been working on a mult-epoch Crab paper which will be expanded to include an \astrosat\ epoch as well. Additionally, due to the recent calibration update in \integral, released in OSA11, a couple of additional epochs of the Crab after the updated calibration will be included as well. Like for G21.5-0.9, the Crab spectrum also has curvature, but at a much higher energy. It is not as abrupt either, but gradually advances from a powerlaw index of 2.1 to 2.14 over the energy range of 80--120~keV. Recent efforts led by E. Jourdain, have shown that a GRBM type of model is a better fit across the 1--150~keV bandpass, and the WG will investigate if the GRBM model should be adopted as a standard.

For other parts of the Crab spectrum, NICER has been working on a Crab model for their calibration, which is using the model from Kaastra et al. (2009) that separates the pulsed spectrum from the PWN. Applying the model from NICER directly to the \nustar\ bandpass resulted in large discrepancies outside of the NICER bandpass, which could be due to the definition of the pulsed spectrum. To investigate this, the WG will start a project to evaluate the pulsed spectrum of the Crab.

\subsection{Thermal SNRs}\label{s:t_snr}\label{s:n132d}

The thermal supernova remnants (SNRs) WG aims to use the time-invariant, line-rich spectra of 1E 0102.2-7219 (E0102), Cas~A, and N132D to improve the response models of the various instruments (gain, CTI correction, QE, spectral redistribution function, etc) and to compare the absolute effective areas of the instruments at the energies of the bright line complexes. Our efforts focus on developing standard models that can be used by the various teams to meaningfully compare their results. The group met remotely five times between September 2020 and August 2021 to discuss the analyses using these standard models.

Andy Beardmore (Leicester) has developed a standard model for Cas~A that he uses for {\it Swift} XRT calibration \footnote{https://wikis.mit.edu/confluence/display/iachec/Cas+A}. Cas~A has the highest flux of any thermal SNR ($\mathrm{F_x}=2.6\times10^{-9}\, \mathrm{ergs~cm^{-2}~s^{-1}}$ [0.3-10.0 keV]) and has strong line emission from Si and S, both of which make it attractive as a calibration source for CCD instruments.  This model provides a good fit to the {\em XMM-Newton} MOS data and is used by
the ASTROSAT team for gain and CTI calibration. 
The {\em Chandra} ACIS calibration team is using this model to fit the  Cas~A data to complement the data acquired from the on-board radioactive source as that source has decayed to less than $0.4\%$ of its flux at launch.

The IACHEC standard model for E0102 \footnote{https://wikis.mit.edu/confluence/display/iachec/Thermal+SNR} is routinely used by several groups for calibration purposes.  E0102 has strong, well-separated lines of O, Ne, \& Mg that have been well-characterized by the RGS on {\em XMM-Newton} (Rasmussen et al. 2001) and the HETG on {\em Chandra} (Flanagan et al. 2004). The model is described in detail in our IACHEC paper Plucinsky et al. 2017.  The model is used regularly by the {\em Chandra} ACIS calibration team to verify updates to the ACIS contamination model (Plucinsky et al. 2020), and it is used by the ASTROSAT and NICER teams for gain and CTI calibration. The E0102 data sets were used as a representative sample in the IACHEC concordance paper (Marshall et al. 2021). Dennerl (MPE) and Plucinsky(SAO) have been using E0102 to test the temperature dependent gain correction for the {\em eROSITA} CCDs.

The majority of the effort of the thermal SNRs WG over the past year has been devoted to the development of a standard model for the LMC SNR N132D.  N132D has the advantage over E0102 that it has significant flux above 2.0~keV including Fe~K emission and it has the advantage over Cas~A that is more compact so that vignetting issues are typically not an issue.  But strong Fe~L emission results in significant line blending with the O and Ne lines in the 0.5-1.0~keV bandpass. The hope of our WG is that E0102 can be the standard thermal SNR source for the 0.5-1.5~keV bandpass and N132D for the 1.5-7.0 keV bandpass.  The early RGS data were published by Behar et al. 2001 and the archival RGS data have been analyzed by Suzuki et al. 2020. The low energy part of the spectrum has been well-characterized by a model with three thermal components.
Stuhlinger(ESAC) used the RGS data to derive an empirical model consisting of Gaussians for the lines and the ``no-line'' APEC model for the continuum components to describe the data.  The high energy part of the spectrum was modeled by Bamba~et al. 2018 based on {\em Suzaku} XIS and {\em NuSTAR} data. They detect Fe~Ly$\alpha$ emission for the first time in N132D and find evidence of a hot plasma that has a temperature of $\sim5$~keV for an equilibrium model or $\sim$1.5~keV for a recombining plasma. 

\begin{figure}[ht]
    \centering
    \includegraphics[width=0.95 \textwidth]{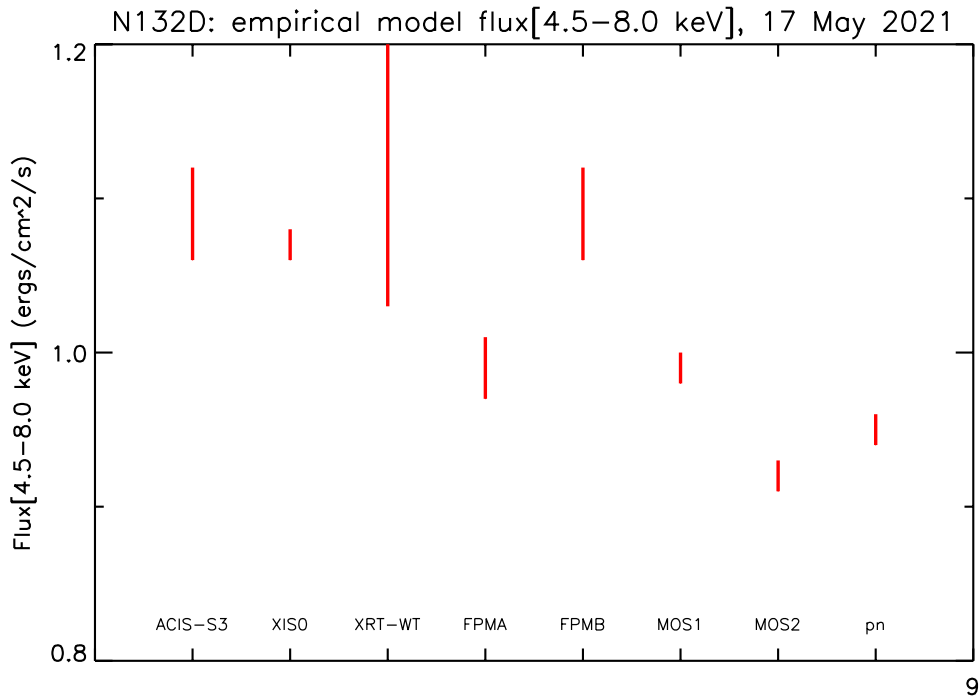}
    \caption{Fluxes in the 4.5-8.0~keV band from N132D fit with the empirical model for ACIS, XIS, XRT, {\em NuSTAR} FPMA/B, MOS, and pn. }
    \label{figure:n132d}
\end{figure}

The thermal SNRs WG has been focused on refining the model for this high temperature component. Grefenstette (Caltech) and Miller (MIT) have combined all available {\em NuSTAR} and {\em Suzaku} data respectively and applied revised background models to constrain this high temperature component. They used these results to construct an ``empirical'' model with a high temperature no-line APEC model with a temperature of 5.47~keV and Gaussians for the Fe lines. Foster (SAO) constructed a ``physical'' model by adopting the Suzuki~et al. (2020) model for the low energy part of the spectrum and adding a {\tt vrnei} model in {\tt XSPEC} with a temperature of 4.7~keV to
model the high energy part of the spectrum.  
The MOS, pn, ACIS, XIS, XRT, \& {\em NuSTAR} data were fit in the 4.5-8.0~keV range with the physical and empirical models by
Foster, Plucinsky, Miller, Beardmore and Grefenstette.  The objective was to freeze most of the parameters and only to allow a few parameters to vary in order to derive a meaningful comparison of the fitted fluxes. For the empirical model only a global normalization, the Fe~XXV~He$\alpha$, and Fe~XXVI~Ly$\alpha$ line complex normalizations were allowed to vary. For the physical model, only the global normalization was allowed to vary.  The flux in the 4.5-8.0~keV band was calculated and compared for the various instruments.  The results are presented in Figure~\ref{figure:n132d}. The ACIS, XIS, and {\em NuSTAR} FPMB fluxes agree to within the uncertainties.  The MOS1, MOS2, pn, and {\em NuSTAR} FPMA fluxes are lower than the ACIS, XIS, and {\em NuSTAR} FPMB fluxes but agree with each other within $1.0\sigma$.  Similar results are achieved with the physical model.

  In the coming year, the thermal SNRs WG plans to finalize the empirical model for N132D in the high energy range (4.5-8.0~keV) and then refit the low energy range (0.5-1.5~keV) with a hopefully simple renormalization.  Once the low and high energy ranges are determined, the group will then fit the 1.5-4.0~keV range in which the CCD instruments have the best combination of collecting area and resolving power. It is clear from our preliminary work that there are significant differences amongst the various instruments in the 0.5-4.0~keV range.  Therefore, no model will provide an acceptable fit to the data from all of the instruments. The WG will have to decide on a reasonable compromise in order to develop a standard spectral model.

\subsection{Timing}\label{s:timing}

 The Timing WG aims to provide a forum for in-orbit and on-ground timing-calibrations of X-ray missions, focusing on their timing systems, calibration methods, issues, and lessons learned. The WG also aims to coordinate simultaneous observations for timing calibrations with multi-X-ray missions and/or radio observatories. Table \ref{table:timing} summarizes the list of in-orbit timing-calibration targets with previous/current missions, which can be useful for future X-ray missions.
 
 As of September 2021, 20 members from 13 missions are participating in the Timing WG, and  we held four virtual local meetings from 2020 to the spring of 2021. Three goals are defined for the WG; i.e., {\bf a)} sharing information of timing calibration methods and protocol, lessens learned to enhance timing capability, {\bf b)} performing the cross-calibration on the timing among multiple missions, analysing systematically the archive data and/or triggering coordinated observations and discussion on the calibration plan for the near future missions, and {\bf c)} studying in detail the effects on the timing products (i.e., light curve, power spectrum, etc) by the detector’s behaviors, such as the dead time, the good-time-interval selection, grade selection, etc.
 
 In 2020 and 2021, the working group members performed two activities related to the goals {\bf a)} and {\bf b)}, which are 1) summary of timing calibration/performance of multiple missions and 2) systematic studies of Crab timing using the archive data among instruments. 

\subsubsection {Summary of timing calibration and performance}
 We gathered the following items from 20 instruments on 12 missions (RXTE, Chandra, XMM-Newton, INTEGRAL, Swift, Suzaku, NuSTAR, Fermi, AstroSat, Hitomi, NICER, and XRISM). The information is summarized on the IACHEC Wiki page\footnote{\it https://wikis.mit.edu/confluence/display/iachec/Timing} in a table with columns:
\begin{itemize}
\item Science Requirement Absolute Time (Requirement \& Goal)
\item Timing System Design (GPS yes/no, Clock Stability)
\item Timing Calibration Status (Timing offset, deviation, \& notes)
\item In-orbit Timing Calibration Targets
\item Reported Issues
\item Reference
\end{itemize}

\begin{figure}[ht]
    \centering
    \includegraphics[width=0.95 \textwidth]{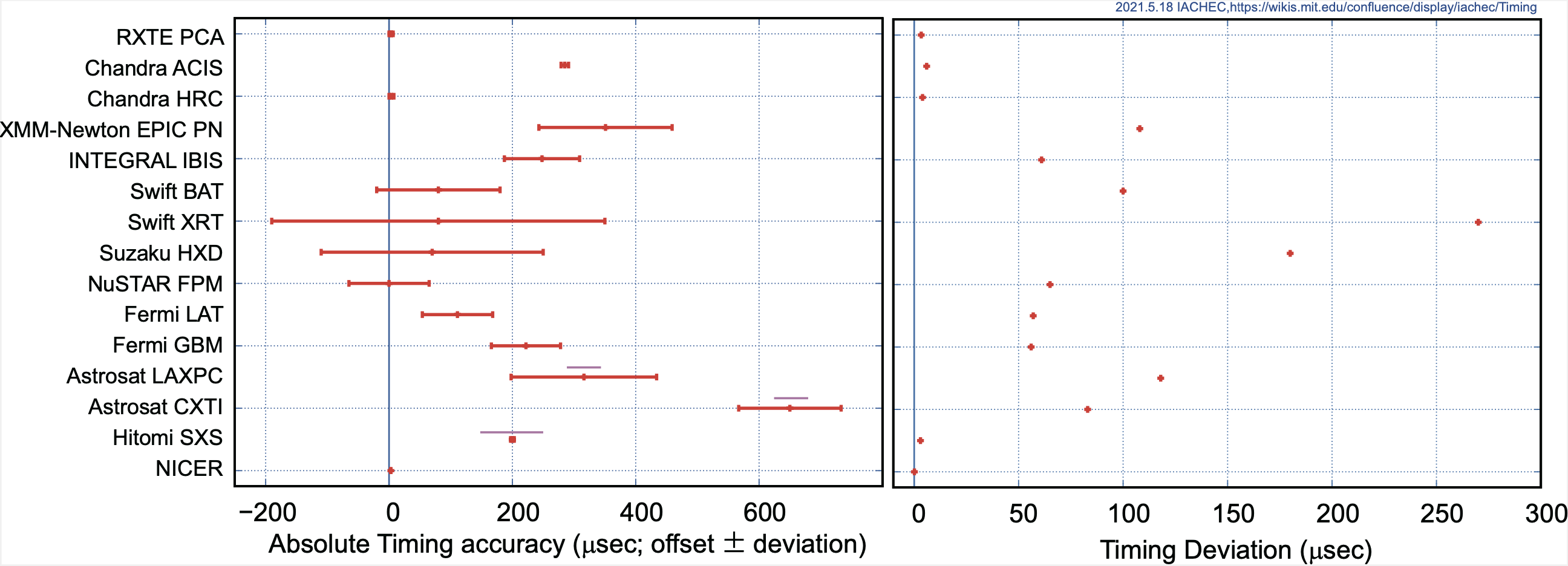}
    \caption{Summary of the timing accuracy on the absolute time and deviation. }
    \label{figure:timing}
\end{figure}

The timing calibration status is summarized in Figure \ref{figure:timing}. Please note that the definition of the offset and deviation in Figure \ref{figure:timing} are not the same among instruments as described in the "Notes" in the "Timing Calibration Status" column of the table at the wiki. In addition, we have to be careful about the energy dependency on the arrival time of the main pulses if the timing accuracy was measured by the Crab observation. 

\begin{table}[h]
    \centering
    \begin{tabular}{c|c}
     \hline 
       In-orbit calibration objects & Mission/Instruments \\
     \hline 
       Crab  & Chandra/HRC, \\
             & XMM-Newtion/EPIC-PN,\\
             & Swift/BAT, Swift/XRT,\\
             & Suzaku HXD,\\
             & NuSTAR/FPM,\\
             & AstroSat/LAXPC, AstroSat/CZTI,\\
             & Hitomi/SXS, Hitomi/HXI, Hitomi/SGD,\\
             & NICER/XTI\\
     \hline 
      PSR B1937+21 & RXTE/PCA, \\
             & Chandra/HRC,\\
             & NuSTAR/FPM,\\
             & NICER/XTI\\
     \hline 
     PSR B1821-24A & NuSTAR/FPM, \\
             & NICER/XTI \\
     \hline 
     PSR B1509-58 & Suzaku/HXD \\
     \hline 
     Gamma-ray bursts & AstroSat/CZTI, Suzaku/HXD-WAM \\
     \hline 
     A0535+262, Her X-1, etc & Suzaku/XIS\\
     \hline 
    \end{tabular}
    \caption{Summary of the Timing calibration targets by previous/current missions.}
    \label{table:timing}
\end{table}

\subsubsection{Systematic timing study with Crab archive data}
To perform a systematic timing-cross-calibration among instruments, we start gathering the barycenter event files of Crab pulsar with multiple missions and check ephemeris of them. 
In April 2021, we checked pulse profiles of Crab with instruments on-board XMM-Newton, Suzaku, NuSTAR, Astrosat, Hitomi, and Swift with an analyses code by M. Bachetti. 
The first quick-look results are shown in the report of the Timing session\footnote{\it https://iachec.org/wp-content/presentations/2021/20210517\_19\_IACHEC\_TimingWGreport.pdf}.
We found several outliers due to ground station issues and are now gathering information on such known issues.

\subsection{Isolated Neutron Stars and White Dwarfs}\label{s:inswd}

This WG aims to improve the cross-calibration of X-ray telescopes in the low energy range ($<$ 1.5 keV) by using spectra of Isolated Neutron Stars (INSs) and White Dwarfs (WDs). These objects should not display time dependent variation and have physically well modeled spectra that can be used as spectral standard candles at low energies. Over the years, a set of white dwarfs (GD153, Sirius B and HZ43) with spectra that can be described by physical white dwarf models, and the isolated neutron star RX\,J1856.5$-$3754 (RX\,J1856) with a spectrum that can be best be described by a single black-body model, have met the characteristics of a standard candle. The three WDs and RX\,J1856  were used to improve the calibration of the low energy end of the \chandra\ LETGS (LETG+HRC-S)and provide a cross-calibration with \rosat\ and recently also with \nicer\  and \erosita. In the context of \erosita\ a new INs is being studied and used for calibration: 1RXS\,J214303.7$+$065419 (RX\,J2143). 

V.~Burwitz gave an overview of the status of the INSs and WDs working group and pointed out that standard RX\,J1856 blackbody model presented on the wiki page needed updating to include the most recent \chandra/LETGS observations. 

H. Marshall presented the status of the \chandra/ACIS calibration using \chandra\ LETG + ACIS observations of RX\,J1856. The results show that the source tracks 44-50 Å (0.248-0.262 keV) throughput very well with the ACIS contaminant reducing the count rate now by factor of 5. The current contaminant model is under-correcting the data in 2017---2019 by 15\,\% compared to 2012---2014. The absolute count rate is 30\,\% less than expected most likely due to PHA losses close to the event threshold. He finds the count rate is vary low, nearly comparable to background  and asks "is it time to increase exposure of the calibration observations?"

V. Burwitz  presented the current status of the \chandra/LETGS observations of RX\,J1856 including the latest June/July 2019 observations that where taken in the context of the \erosita\ in-flight calibration. All \chandra/LETGS observations of RX\,J1856 span a period of about 20 years. The main result is that spectrum remains constant over the 20 years and does not change shape. Therefore it can continue to be used as as standard candle. 

F. Haberl presented the results obtained of observations of ROSITA calibration observations of the INSs RX,J1856 and RXJ\,J2143. The observations are used to monitor the potential buildup contamination during the mission especially after larger orbital manoeuvres. The good news is that so far no contamination buildup has been detected. But compared to ground based calibration a small C-K absorption edge is required to be able to fit the RX\,J1856 standard single black-body model to the data. New INs to using for \erosita\, calibration is RXJ\,J2143. I has higher temperature ($\sim$100\,eV) black-body spectrum, but it require a further spectral component to best fit the \erosita\ spectra.  Whilst fitting spectra using single events provides highest spectral resolution at the cost of having different fluxes varies for different Telescope Module (seven) and observations. This is caused by a different location of the peak of the PSF on the 75\,µm$\times$75\,µm pixels. The sub-pixel position of the source effect will have to be studied and further calibrated. But using all valid events (single, double, triple and quadruple events) provides consistent fluxes between telescope modules and individual observations.

For RX\,J1856 a question was raised about the detection in \xmm\  and \chandra/ACIS data and whether this can be confirmed with \erosita\ observations, this is being looked into.

In summary, the INS RX\,J1856.5$-$3754 is being observed and used by all X-ray observatories for calibration purposes to monitor the status of their low energy calibration. Work on WDs was not discussed at this year's meeting.

\section*{References\footnote{see {\tt https://iachec.org/papers/} for a complete list of IACHEC papers}}

\noindent
Algeri, S., 2020, Phys.Rev.D, 101, 015003, \href{https://arXiv.org/abs/1906.06615}{arXiv:1906.06615} \\
\noindent
Bamba et al.\ 2018, ApJ, 854, 71\\
\noindent
Bartalucci, I., Mazzotta, P., Bourdin, H., and Vikhlinin, A., 2014, A\&A, 566, A25\\
\noindent
Behar, E., et al. 2001, A\&A, 365, L242\\
\noindent
Belanger, G., 2013, ApJ, 773, 66 \href{https://arxiv.org/pdf/1303.7408.pdf}{ApJ, 773, 66}\\
\noindent
Belanger, G., 2016, ApJ, 822, 14 \href{https://arxiv.org/pdf/1712.00734.pdf}{ApJ, 822, 14}\\
\noindent
Chen Y., et al., 2019, \href{https://doi.org/10.1080/01621459.2018.1528978}{J.Am.Stat.Assoc., 114:527, 1018}\\ 
\noindent
Drake, J.J., et al. 2006, Proc. SPIE, 6270, 62701I\\
\noindent
Flangan, K.A., et al. 2004, ApJ, 605, 230\\
\noindent
Gu, L., et al. 2020, A\&A, 614, 93\\
\noindent
Hitomi Collaboration et al. 2018, PASJ, 70, 16\\
\noindent
Ishida M., et al.\ 2011, PASJ, 63, 657\\
\noindent
Kaastra J., et al, 2009, A\&A, 497, 291\\
\noindent
Kaastra J. \& Bleeker J.A.M., 2016, A\&A, 587, 151 \\
\noindent
Kaastra, J.S.\ 2017, A\&A, 605, A51\\
\noindent
Kashyap, V.L., 2020, {\sl Basics of Astrostatistics}, in Tutorial Guide to X-ray and Gamma-ray Astronomy, p.203, Ed.\ C.Bambi, Springer, ISBN 978-981-15-6337-9 \\
\noindent
Kettula, K., Nevalainen, J., \& Miller, E.D.\ 2013, A\&A, 552, 47\\
\noindent
Koyama, K., et al.\ 2007, PASJ, 59, 23\\
\noindent
Lee, H., et al.\ 2011, ApJ, 794, 97\\
\noindent
Madsen K., et al.\ 2017a, AJ, 153, 2 \\
\noindent
Madsen K., et al.\ 2017b, AJ, 841, 5 \\
\noindent
Marshall H., et al.\ 2004, Proc.\ SPIE, 5165, 497 \\
\noindent
Marshall, H., 2021a, AJ, 162, 134 \\
\noindent
Marshall, H., 2021b, AJ, 907, 82 \\
\noindent
Marshall H., et al.\ 2021, AJ, \href{https://arxiv.org/pdf/2108.13476.pdf}{arxiv.org:2108.13476} \\
\noindent
Nevalainen, J., David, L., \& Guainazzi, M.\ 2010, A\&A, 523, A22\\
\noindent
Nevalainen, J., Valtchahov, I., Saxton, R.D., \& Molendi, S.\ 2021, XMM-Newton calibration technical note XMM-SOC-CAL-TN-0227 (\href{https://arxiv.org/abs/2103.01753}{arXiv:2103.01753})\\
\noindent
O'Dell S., et al.\ 2013, Proc.\ SPIE, 8559 \\
\noindent
Plucinsky P., et al.\ 2016, Proc.\ SPIE, 9905, 44 \\
\noindent
Plucinsky P., et al.\ 2017, A\&A, 597, A35 \\
\noindent
Plucinsky P., et al.\ 2020, Proc.\ SPIE, 11444, 1144497\\
\noindent
Rasmussen, A.P, et al.\ 2001, A\&A, 365, L231\\ 
\noindent
Rasmussen, C.E.\ and Williams, C.K.I. 2006, {\sl \href{http://www.gaussianprocess.org/gpml/chapters/}{Gaussian Processes for Machine Learning}}, MIT Press, ISBN 026218253X\\
\noindent
Read, A.M., Guainazzi, M., and Sembay, S., 2014, A\&A, 564, A75\\
\noindent
Reiprich, T.H. \& B\"ohringer, H.\ 2002, ApJ, 567, 2, 716\\
\noindent
Schellenberger, G., et al.\ 2015, A\&A, 575, 30\\
\noindent
Stuhlinger, M., et al.\ 2010, \href{http://xmm2.esac.esa.int/docs/documents/CAL-TN-0052.ps.gz}{XMM-SOC-CAL-TN-0052 6.0}\\
\noindent
Suzuki, H., et al.\ 2020, ApJ. 900, 39\\
\noindent
Tanaka, T., et al.\ 2018, JATIS, 4, 011211\\
\noindent
Tashiro, M., et al.\ 2018, Proc.\ SPIE, 10699, 1069922\\
\noindent
Terada Y., et al., 2008, PASJ, 60, 25\\
\noindent
Terada, Y., et al., 2018, \href{https://doi.org/10.1117/1.JATIS.4.1.011206}{J.Astron.Tel.Instr.Sys.\ (JATIS), 4(1), 011206}\\ 
\noindent
Terada Y., et al., 2018, JATIS, 011206\\
\noindent
M.Tsujimoto, et al., 2011, A\&A, 525, 25
\noindent
Watson, M.G., et al., 2009, A\&A, 493, 339\\
\noindent
Xu, J., et al.\ 2014, ApJ, 794, 97\\
\noindent
Yu, X., et al.\ 2018, ApJ, 866, 146\\

\end{document}